\documentclass[12pt]{iopart}

\expandafter\let\csname equation*\endcsname\relax
\expandafter\let\csname endequation*\endcsname\relax

\usepackage{amsmath}
\usepackage{amssymb}
\usepackage{array}
\usepackage[hidelinks]{hyperref}
\usepackage{stfloats}
\usepackage{url}
\usepackage{graphicx}
\usepackage{caption}
\usepackage{subcaption}
\usepackage{cite}
\usepackage{xcolor}

\newcommand{\dkl}{D_{\text{KL}}}

\newcommand{\E}{\mathbb{E}}
\newcommand{\lp}{\log\!\text{-}\mathrm{cap}}

\newcommand{\p}{\mathrm{p}}
\newcommand{\R}{\mathrm{r}}
\newcommand{\B}{\mathrm{b}}
\newcommand{\odds}{\mathrm{o}}

\newcommand{\diff}{\mathrm{d}}

\newcommand{\prob}[1]{\mathbb{P} \left[ #1 \right]}
\newcommand{\condprob}[2]{\mathbb{P} \left[ \left. #1 \right| #2 \right]}
\newcommand{\condprobs}[2]{\mathbb{P} \left[ #1 \left| #2  \right. \right]}

\renewcommand{\underline}[1]{\mathbf{#1}}
\newcommand{\intsimpldetail}[2]{\int_{\mathbb{R}^M_+}\diff \mathbf{b} ~ \delta \left( \sum_{\tilde{x}} \B_{\tilde{x}} - #1 \right) ~ #2}

\begin{document}

\title[Adaptive strategies in Kelly's horse races model]{Adaptive Strategies in Kelly's Horse Races Model}
\author{Armand Despons$^1$, Luca Peliti$^2$ and David Lacoste$^1$}
\address{$^1$ Gulliver Laboratory, UMR CNRS 7083, PSL Research University, 
ESPCI, 10 rue Vauquelin, F-75231 Paris Cedex 05, France}
\address{$^2$ SMRI, Santa Marinella (RM), 00058 Italy}

\ead{\href{mailto:armand.despons@outlook.fr}{armand.despons@espci.fr}}
\begin{indented}
	\item[]May 2022
\end{indented}

\begin{abstract}
	We formulate an adaptive version of Kelly's horse model in which the gambler learns from past race results using Bayesian inference. We characterize the cost of this gambling strategy and we analyze the asymptotic scaling of the difference between the growth rate of the gambler and the optimal growth rate, known as the gambler's	regret. We also explain how this adaptive strategy relates to the universal portfolio strategy, and we build improved adaptive strategies in which the gambler exploits the information contained in the bookmaker odds distribution.
\end{abstract}	


\maketitle
\tableofcontents
\pagestyle{empty}

\section{Introduction}

In 1956, R. Kelly \cite{kelly_new_1956}  extended the work of C. Shannon on communication to gambling models. This now classic piece of information theory \cite{cover_elements_2006} turned out to have fundamental implications for investment strategies in finance and beyond. In the context of biology, Kelly's paper lead to a clarification of the notion of value of information \cite{rivoire_value_2011}. This value of information quantifies the quality of the strategies adopted by individuals in a population when faced with unpredictably varying environments \cite{bergstrom_shannon_2004}. In these conditions, it is sometimes advantageous for individuals to accept a reduction of their short-term reproductive success, in exchange for longer-term risk reduction, a strategy known as  \emph{bet-hedging} \cite{levien_non-genetic_2021}. Such a strategy is employed for instance by cells to cope with antibiotics or by plants to cope with a fluctuating climate \cite{venable_bet_2007}. 

Kelly's strategy is known to be risky, and for this reason most gamblers use fractional Kelly's strategies, with reduced risk and growth rate \cite{maclean_kelly_2011}. This observation hints at a trade-off between the risk the gambler is ready to take and the average long-term growth rate of his capital. In a recent work, we have studied the properties of this trade-off inspired by ideas in Stochastic Thermodynamics \cite{dinis_phase_2020}. 
We then studied another manifestation of this trade-off this time in the context of a biological population with phenotypic switching in a fluctuating environment \cite{dinis_pareto_2022}. 

In that work, the phenotypic switching rate of individuals in the population is not affected by the state of the environment, which means that no sensing mechanism is present.
Naturally, real biological systems are able to sense their environment, extract relevant information from noisy measurements, and adjust their phenotypic response accordingly \cite{kussell_phenotypic_2005}. An open question in this field is thus the precise quantification of the cost that biological systems must pay for this sensing.

This question motivates the present paper. To evaluate this cost, 
we go back to Kelly's model, and we modify it to add some adaptation mechanisms based on Bayesian inference \cite{perkins_strategies_2009}. 
The adaptive strategy we build in that way has similarities with the one used in universal portfolios \cite{cover_universal_1991} and is therefore optimal among all the strategies of this type. 
We quantify the cost of our adaptive strategy, which in the context of Kelly's horse races takes the form of a capital loss of the gambler with respect to his/her optimal capital. We derive an asymptotic form for this cost both for uncorrelated and correlated races. We then investigate improved strategies in which the gambler leverages the information contained in the odd distribution chosen by the bookmaker. 

\section{Definition of Kelly's model}

Let us recall the basic elements of Kelly's horse race model \cite{kelly_new_1956}. 
A race involves $M$ horses, and is described by a normalized vector of winning probabilities $\mathbf{p}$, an inverse-odds vector $\mathbf{r}$ and a gambler strategy $\mathbf{b}$. 
The latter corresponds to a specific allocation of the gambler's capital on the $M$ horses: if we denote by $C_t$ the gambler's capital at time $t$, the amount of capital invested on horse $x$ reads $ \B_x C_t $. 
We further assume that, after each race, the gambler invests his whole capital, i.e., $\sum_{x=1}^{M} \B_x = 1$, always betting a non-zero amount on all horses, i.e., $\forall x \in [1 , M]:\  \B_x \neq 0$. 
The inverse-odds vector $\mathbf{r}$ is set by the bookmaker and is not necessarily normalized, in fact the value of $\sum_x \R_x$ controls whether or not the bookmaker is extracting/injecting 
additional capital: if the sum of the components of the inverse-odds is strictly bigger (resp. strictly lower) than 1 the bookmaker extracts (resp. injects) additional capital. 
In the following, we focus on the neutral case where the bookmaker neither extracts nor injects capital, which means that the inverse-odds vector is normalized: $\sum_x \R_x = 1$. 
This case is usually referred to as gambling without \emph{track-take} in the horse-racing literature \cite{kelly_new_1956, cover_elements_2006}. 

Thus, the evolution of the gambler's capital after one race reads: 
\begin{equation}
	\label{cap_evol}
	C_{t+1} = \dfrac{\B_x}{\R_x} C_t ,\quad \text{with a probability } \p_x,
\end{equation}
which implies that the log of the capital, $\lp (t) \equiv \log C_t$, evolves additively:
\begin{equation}
	\label{eq:log_cap_evol}
	\lp (t) = \sum_{\tau = 1}^t \log \left(  \dfrac{\B_{x_\tau}}{\R_{x_\tau}} \right) ,
\end{equation}
where $x_\tau$ denotes the index of the winner of the $\tau$-th race and we assumed $\lp(0) = 0$ (i.e. $C_0 = 1$).
Since the races are assumed to be independent, the terms $\log (\B_{x_\tau}/ \R_{x_\tau})$ in \eqref{eq:log_cap_evol} are independent and identically distributed, and we can use 
the weak law of the large numbers : 
\begin{align}
	\dfrac{\lp (t)}{t} \xrightarrow[t \to \infty]{} \mathbb{E} \left[ \log \left(  \dfrac{\B_{x}}{\R_{x}} \right)  \right] 	
	\label{long_term_growh}
\end{align}
in probability. Then 
\begin{equation}
	\label{eq:avg_capital_increases}
	\mathbb{E} \left[ \log \left(  \dfrac{\B_{x}}{\R_{x}} \right)  \right] \equiv \sum_x \p_x \log \left(  \dfrac{\B_{x}}{\R_{x}} \right) = D_{KL} \left( \bold{p} || \bold{r} \right) - D_{KL} \left( \bold{p} || \bold{b} \right),
\end{equation}
where $\dkl$ stands for the Kullback-Leibler divergence \cite[sec.~2.3]{cover_elements_2006}.
From an information theoretic point of view, \eqref{long_term_growh} and \eqref{eq:avg_capital_increases} imply that the capital of the gambler increases in the long term 
only if the gambler has a better knowledge of $\mathbf{p}$ than the bookmaker, otherwise it decreases. 

It also follows from this analysis that the optimal strategy $\mathbf{b}^\text{\tiny{KELLY}} = \mathbf{p}$, called 
Kelly's strategy \cite{kelly_new_1956}, overtakes any other strategies in the long-term. 
Its optimum growth rate is the positive quantity 
$\dkl \left( \mathbf{p} \Vert \mathbf{r} \right)$. 

Note that thanks to the assumption that the inverse-odds vector is normalized, another strategy is possible, namely $\mathbf{b}^{\text{\tiny NULL}} = \mathbf{r} $. 
We have called this strategy the \emph{null strategy} \cite{dinis_phase_2020}, because it yields a constant capital as can be seen from 
Eqs. (\ref{long_term_growh})-(\ref{eq:avg_capital_increases}). Indeed, in this null strategy, 
the gambler has exactly the same knowledge as the bookmaker, hence he/her neither looses nor gains capital. 

\section{Adaptive strategy}

Kelly's strategy assumes that the probability vector $\underline{p}$ is known, but this, in practice, is never the case. 
Hence, it is of great interest to build strategies to learn $\mathbf{p}$, 
in order to play the optimal betting strategy. In information theory, these strategies are called universal precisely 
because they do not require any knowledge of $\mathbf{p}$. From a theoretical point of view,
universal strategies are related to lossless source coding~\cite{merhav_universal_1998}.
 
While general theories of history-dependent gambling are available 
\cite{hirono_jarzynski-type_2015}, here we want to construct an adaptive strategy specifically for Kelly's problem. 
We use an estimator based on \emph{Laplace's rule of succession}  \cite{jaynes_probability_2003} :
\begin{equation}
	\label{laplace_bet}
	\B_x^{\text{\tiny{LAPL}},~t+1}  \equiv \dfrac{n_x^t + 1}{t+M},
\end{equation}
where the vector $\mathbf{n}^t$ contains the history of all race results up to time $t$: $ \underline{n}^t \equiv \left[ n_1^t, \hdots, n_M^t \right] ^\mathrm{T} $. In this notation, $n_x^t$ represents the number of times the horse $x$ has won among the $t$ previous races, and obviously $\sum_{x=1}^{M} n_x^t = t$. 

\begin{figure}[t]
	\centering
	\begin{subfigure}[b]{0.495\textwidth}
		\centering
		\includegraphics[scale=1]{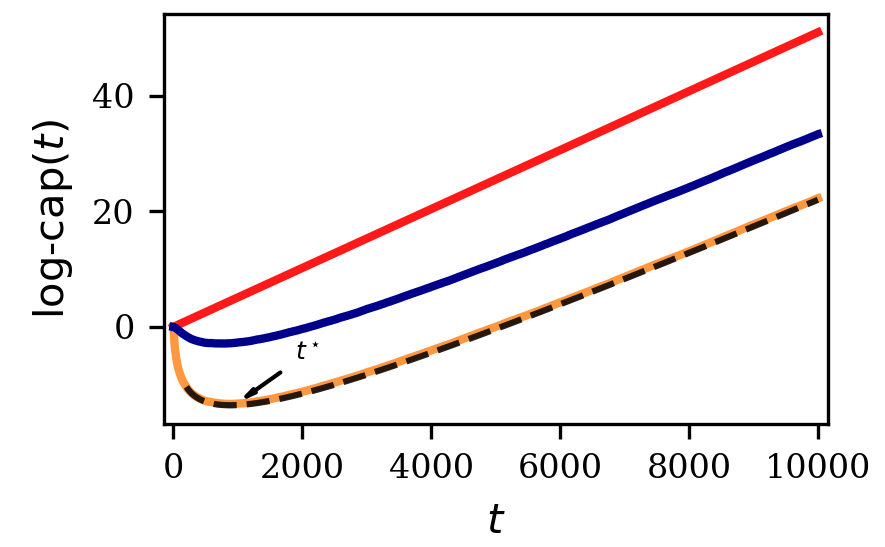}		
		\caption{\label{fig}}
	\end{subfigure}
	\begin{subfigure}[b]{0.495\textwidth}
		\centering
		\includegraphics[scale=1]{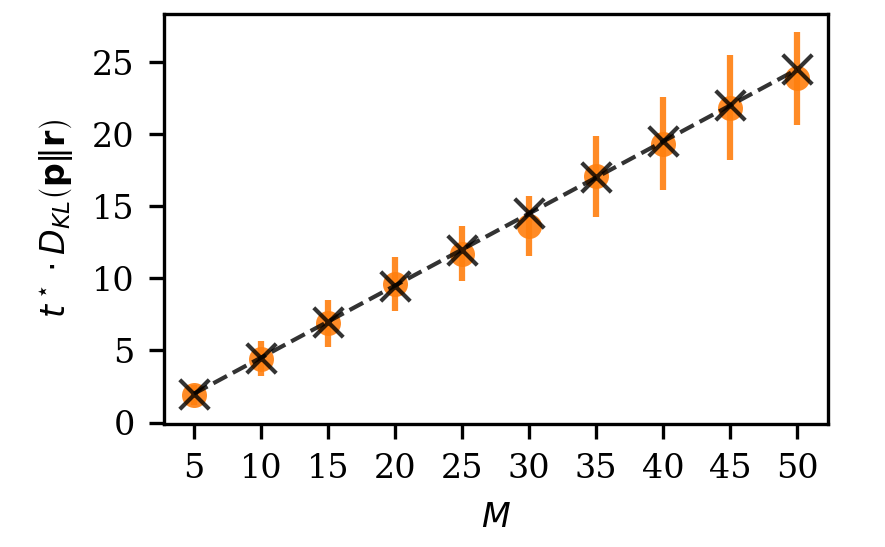}
		\caption{\label{fig2}}
	\end{subfigure}	
	\caption{(a) Time evolution of the log-capital of Kelly's strategy (red solid curve), of Laplace's strategy \eqref{laplace_bet} (orange solid curve) and of the modified Laplace's strategy using the odds distribution
	as prior \eqref{eq:ps_lapl_hypothesis} (blue solid curve).
	Both adaptive strategies undergo a \emph{learning phase} in which the log-capital decreases during a time $t^\star$, and in the long run they perform as well as Kelly's strategy. 
	Note that the modified Laplace's strategy reduces the initial loss of capital associated to the learning phase.   
	The average log-capital associated with Laplace's strategy \eqref{asympt_adapt_log_cap} is the black dotted curve.
	In this figure, the log-capital is averaged over 3000 realizations with $M=10$ and $\epsilon =0.1$. 
	(b) Linear dependence of the rescaled $t^\star$ 
	as function of $M$, the number of horses. 
	The color dots represent numerical estimates with error bars. The black crosses and the dashed curve represent the analytical prediction of \eqref{t_opt_expr}. 
	For the each colored dot, we have chosen $\mathbf{p}$ and $\mathbf{r}$ randomly for a fixed $M$ and $\epsilon = 0.1$ and we have 
	averaged over 50 realizations of $t^\star$, itself computed from the average over 30 realizations of the log-capital.
	}
\end{figure}
	
This result \eqref{laplace_bet} follows from a standard application of Bayesian inference : 
one starts from the likelihood of $\mathbf{n}^t$, which is the multinominal distribution:
\begin{equation}
	\label{eq:likelihood}
	\prob{\left. \mathbf{n}^t \right| \underline{b} } = \dfrac{t!}{n_1^t ! \cdots n_M^t !} \B_1^{n_1^t} \cdots \B_M ^{n_M^t}.
\end{equation}
The posterior distribution is then given by \emph{Bayes' rule}: $\condprobs{ \underline{b} }{ \mathbf{n}^t } = \dfrac{\condprob{ \mathbf{n}^t }{\underline{b}} \prob{ \underline{b} }}{\prob{\mathbf{n}^t }}$. Assuming a uniform prior $\prob{\mathbf{b}} \propto \delta \left(   \sum_x \B_x -1 \right)$,  
the posterior distribution reads: 
\begin{equation}
	\label{eq:posterior_distrib}
	\condprobs{\mathbf{b}}{\mathbf{n}^t}= \dfrac{(t+M-1)!}{n_1^t ! \cdots n_M^t !} ~ \B_1^{n_1^t} \cdots \B_M ^{n_M^t} ~ \delta \left( \sum_x \B_x - 1 \right).
\end{equation}
Integrating \eqref{eq:posterior_distrib} with respect to all the components except one yields :
\begin{equation}
	\label{eq:prob_one_compo_normalized}
	\condprob{\B_x}{\mathbf{n}^t} = \dfrac{(t+M-1)!}{n_x^t ! ~ (t - n_x^t + M - 2) ! } \B_x^{n_x^t} (1-\B_x)^{t - n_x^t + M - 2}.
\end{equation}
\emph{Laplace's rule of succession} \eqref{laplace_bet} is then obtained by taking the expectation value of the posterior 
for each component $\B_x$. 
Alternate estimators are acceptable such as the one based on the maximum of the posterior, but typically 
the Laplace estimator performs better specially when one or several horses have a very low probability of winning. 
In fact, the form of Eq.~\eqref{laplace_bet} guarantees that all components of the vector of estimated bets are strictly positive even if the conditional probability given by Eq.~\eqref{eq:prob_one_compo_normalized} may become zero.
In particular at $t=0$, the Laplace estimator is $\B_x=1/M$ which means that the gambler 
divides his capital uniformly among the $M$ horses, clearly the 
best thing to do in the absence of any information prior to the first race.
The asymptotic regime of $\B_x^{t}$ at large times $t$ can be analyzed using the central limit theorem. 
Only at large times does the estimator based on Laplace's rule of succession become unbiased. In Appendix A \eqref{eq:dkl_theory}, we show that the KL divergence between $\mathbf{p}$ and $\mathbf{b}^{\text{LAPL},~t}$ behaves as:
\begin{equation}
	\label{asympt_dkl}
	\dkl \left( \underline{p} \Vert \mathbf{b}^{\text{\tiny{LAPL}},~t} \right)   \underset{t \to \infty}{\sim} \dfrac{M-1}{2 t} \xrightarrow[t \to \infty]{} 0.
\end{equation}
The same calculation also leads to an explicit expression of the standard deviation of the Laplace estimator. We obtain:
\begin{equation}
	\label{eq:std_estimator}
	\mathrm{STD} \left[ \B_x \vert \mathbf{n}^t \right] = \sqrt{\dfrac{\B_x^{\text{\tiny{LAPL}}, ~t}(1-\B_x^{\text{\tiny{LAPL}}, ~t})}{t + M +1}}.
\end{equation}

Here we assumed that the distribution $\mathbf{p}$ is constant in time. 
Laplace's strategy would still perform well if $\mathbf{p}$ was a slowly varying function of times, i.e. if that distribution was varying on a 
time scale much longer than that of $\mathbf{b}^{\text{\tiny LAPL}}$, which by \eqref{asympt_dkl} is given by $(M-1) /2$.

\subsection{Optimality of the strategy}

As mentioned in the introduction, Laplace's rule qualifies as universal because no foreknowledge 
of the distribution of the races $\underline{p}$ is needed, instead the distribution is learned over time. In the literature on portfolios, 
a strategy of this type is known under the name of \textit{universal portfolio}. This strategy is important in practice because it 
performs asymptotically as well as the best constant balanced portfolios which would have a foreknowledge of the stock prices \cite{cover_universal_1991}. 
Here, we explain precisely in what sense these two strategies are related. 
In the portfolio case, $\mathbf{b}$ denotes the proportion of wealth invested on a certain day in an ensemble of stocks. The total capital is multiplied after each day by the product of $\mathbf{b}$ with a matrix which describes the increase in the price of all the stocks from one day to the next. Thus, in the portfolio case, the gain of the capital is distributed on all the selected stocks, in contrast to Kelly's model in which a single horse is selected at each race.

To connect this strategy to Kelly's horse race, let us consider this portfolio in the case where only a single stock contributes each day. 
Let us then denote the index of that stock up to time a $t$ in a vector $\mathbf{x}^t \equiv \left[ x_1, \cdots, x_t \right]^\text{T}$, and let $C \left[ \mathbf{b}, t  \vert \mathbf{x}^t \right] $ be the total capital a portfolio $\mathbf{b}$ would have gained 
given the history $\mathbf{x}^t$ up to time~$t$. According to Ref.~\cite{cover_universal_1991}, the bet component $\B_{x}$ in the next race is given by
\begin{equation}
	\label{eq:univ_estimator_general}
	\B_{x}^{\text{\tiny PORTF},~t+1} \equiv \dfrac{ \intsimpldetail{1}{\B_x} ~ C \left[ \mathbf{b}, t  \vert \mathbf{x}^t \right] }{\intsimpldetail{1}{C \left[ \mathbf{b}, t  \vert \mathbf{x}^t \right]}}.
\end{equation}
In the case of Kelly's model, where $C \left[ \mathbf{b}, t \vert \mathbf{x}^t \right] = \prod_{i=1}^{t} \B_{x_i} \odds_{x_i} $, the integrals above can 
be performed explicitly and one recovers exactly \eqref{laplace_bet}. 
Therefore, the adaptive strategy based on Laplace's rule of succession outlined above and the universal portfolio strategy are equivalent when considering a single stock. Since the universal portfolio is known to outperform any other strategy asymptotically, it follows that the adaptive strategy based on Laplace's rule is also optimal among all strategies given the assumptions of the model. 
Naturally, the universal portfolio has a wider applicability and is more suited to the stock market, because it accounts for distributed gains among many stocks. 
In this case of distributed gains, interesting and far reaching connections between Kelly's horse race and optimality in game theory have been studied  \cite{pugatch_asymptotic_2014}.

\subsection{The cost of learning} 

A typical evolution of the log-capital in the adaptive strategy is shown in Figure \ref{fig} (orange curve), together with the optimal strategy based on Kelly's criterion (red solid curve).  
The figure shows that the gambler loses capital at short times because of his/her poor knowledge of the distribution $\underline{p}$, but he/she catches up on long times once the distribution has been learned.

The duration of the burn-in phase is an interesting quantity, which we call $t^*$ and which is shown in Figure~\ref{fig}. 
This time is a proxy for the actual loss of capital during the burn-in phase, and for this reason we may regard it as the cost of learning. 
Clearly, this time depends on the \emph{knowledge} of the bookmaker : 
if the bookmaker has a perfect knowledge of the races, he can set the odds according to ($\underline{r} = \underline{p}$), in which case 
the burn-in time is very large and the log-capital will only decrease.  
On the contrary, if the bookmaker has a poor knowledge of the system, the inverse odds are far from $\underline{p}$, and in 
that case, the burn-in time tends to be very short. 
In practice, we introduce a parameter $\epsilon$ and a random vector $\boldsymbol{\eta}$ uniformly distributed in the simplex with 
$\mathbf{r} = \mathbf{p} + \epsilon ~ (\boldsymbol{\eta} - \mathbf{p} )$, such that for $\epsilon = 0$ 
one has $\mathbf{r} = \mathbf{p}$ and for $\epsilon=1$, $\mathbf{r}$ is independent of $\mathbf{p}$. 

To quantify the effect more precisely, it is convenient to define $t^\star$ as the minimum of the log-capital in the adaptive strategy, and to focus on the difference of log-capital between Kelly's strategy and  Laplace's strategy: $\label{log_cap_decomp_delta}
\Delta (t) \equiv   \lp^\text{KELLY} (t) - \lp (t) $.
The quantity $\Delta (t) $ has been considered before in the multi-armed bandit problem \cite{lai_asymptotically_1985} or in the portfolio theory \cite{cover_universal_1991, cover_universal_1996}, it is usually called the \emph{regret} associated with the adaptive strategy.  
The important point here is that the regret is odd-independent: 
$ \Delta (t) = \left. \left. \sum_{i=1}^{t} \right[  \log \p_{x_i} - \log \B_{x_i}^{\text{LAPL}, ~ i} \right] $. 

Averaging the regret associated with Laplace's strategy on the race realizations yields (see Appendix A for details on the derivation):
\begin{equation}
	\label{avg_delta_expr}
	\left\langle  \Delta (t) \right\rangle  = \left\langle  \Delta (t_0) \right\rangle   +\dfrac{M-1}{2} \log \dfrac{t}{t_0+1},
\end{equation}
for $t_0 > \underset{x}{\max}~\p_x ^{-1}$.
From \eqref{avg_delta_expr} one sees that $\left\langle \Delta (t) \right\rangle \sim \log t$, a feature  already observed for the optimal adaptive strategy in the multi-armed bandit problem \cite{lai_asymptotically_1985}.  
Since we know the long-term behavior of Kelly's strategy from \eqref{long_term_growh}, we deduce an expression for the average log-capital of our adaptive strategy:
\begin{equation}
	\label{asympt_adapt_log_cap}
	\left\langle \lp (t) \right\rangle = \dkl \left( \underline{p} \Vert \underline{r} \right) t - \dfrac{M-1}{2} \log \dfrac{t}{t_0 + 1} -  \left\langle  \Delta (t_0) \right\rangle .
\end{equation}
This theoretical prediction is shown in Figure \ref{fig} (black dotted curve) for a time $t_0$ respecting the condition for \eqref{avg_delta_expr}.  
By taking the derivative of \eqref{asympt_adapt_log_cap} with respect to time, one obtains the following estimate for $t^\star$:
\begin{equation}
	\label{t_opt_expr}
	t^\star = \dfrac{M-1}{2}\dfrac{1}{ \dkl \left(  \underline{p} \Vert \underline{r} \right) }.
\end{equation}
As shown in Figure \ref{fig2}, this time $t^\star$ indeed  increases linearly with $M$ and is inversely proportional to the KL-divergence between $ \underline{p}$
and $\underline{r}$, which quantifies the knowledge of the races by the bookmaker.
Since \eqref{t_opt_expr} is based on an asymptotic result, namely \eqref{asympt_adapt_log_cap}, this expression over-estimates the real $t^\star$ 
but becomes accurate in the limit when $t^\star$ is larger than the time $t_0$.

At this point, it is important to appreciate that $t^\star$ in the above formula still depends on $\mathbf{p}$, 
which is the very quantity we are trying to learn in the adaptive strategy, therefore 
\eqref{t_opt_expr} could appear useless. In fact, this is not so, because the gambler can still use $\mathbf{b}^{\text{LAPL},~t}$ in \eqref{t_opt_expr} 
to obtain an estimate for the time  $t^\star$. In practice, if the gambler is allowed not to play initially, he/she
should wait this time $t^\star$ before starting to bet.

\subsection{Extension to correlated races}

It is straightforward to extend the above strategy to the case of correlated races, provided they are Markovian, 
in other words, when the results of each race only depends on the result of the previous race. 
In such an extension, the quantity to be learned is no longer a vector of $M$ components, but an 
$M \times M$ matrix $\left\lbrace \mathrm{p}_{x,y} \right\rbrace$ whose columns $\mathbf{p}_{\odot \vert y }$ are the conditional probability 
of winning given that horse $y$ has won in the previous race. Similarly, the odds and the bets are also matrices of size $M \times M$.

Let us assume that the associated Markov chain is irreducible, and thus admits a unique stationary distribution $\underline{p}^{\text{s}}$. 
The equation governing the evolution of the log-capital is
\begin{equation}
	\label{log_cap_evol_markov}
	\lp (t)  \underset{t \to \infty}{\sim} \left(  \sum_{x,y=1}^M \p_{x,y} \log \dfrac{\B_{x|y}}{\R_{x|y}} \right) t,
\end{equation}
where $\p_{x,y} \equiv \p_{x|y} \cdot \p^\text{s}_{y}$ and where $\R_{x|y}$ are the inverse 
odds conditioned on the last race's result.
Kelly's strategy is again the one that maximizes the long-term growth rate, given the normalization of the conditional bets 
$\mathrm{b}_{x,y}^\text{KELLY} = \mathrm{p}_{x,y}$ \cite{cover_elements_2006}.

We now adapt Laplace  rule's of succession to take into account correlations:
\begin{equation}
	\B_{x|y}^{\text{LAPL},~t+1} = \dfrac{n_{x|y}^t + 1}{\tau_{y}^t + M},
\end{equation}
where $n_{x|y}^t$ and $\tau_y^t$ are respectively the number of times horse $x$ wins immediately following a victory of horse $y$, and the number of times horse $y$ 
has won irrespective of the previous events. 

Qualitatively the behavior for correlated or uncorrelated races are similar, 
except that the data to be learned in the correlated case lies in a higher dimensional space, and as a result, the learning phase lasts longer 
and the losses in terms of log-capital are heavier.
Indeed, the regret now scales as $M^2$ while still being logarithmic in time (see Appendix C):
\begin{equation}
	\left\langle \Delta (t) \right\rangle = \left\langle \Delta (t_0) \right\rangle  + \dfrac{M(M-1)}{2} \log \dfrac{t}{t_0 + 1}.
\end{equation}
In the end, one obtains the following estimate of the learning time :
\begin{equation}
	\label{t_opt_expr_markov}
	t^\star = \dfrac{M(M-1)}{2 \left\langle \log \dfrac{\p_{x|y}}{\R_{x|y}} \right\rangle}.
\end{equation}
In fact one could have predicted the behavior in $M(M-1)$ because in the learning of the $M \times M$ matrix, the last row is determined by the constraint on the normalization.

The extension to Markov races of higher order $n$ is straightforward.
Now, the relevant distribution is $\p_{x| \mathbf{Y}}$
where $\mathbf{Y}$ store the $n$ previous winners.
Thus, the matrix of the conditional probability becomes a $M \times M^n$ matrix. 
As before, the last row is determined by the normalization of each conditional probability, hence the effective number of degrees of freedom is $M^n(M-1)$. 
Therefore, one obtains:
\begin{equation}
	\left\langle \Delta (t) \right\rangle = \left\langle \Delta (t_0) \right\rangle  + \dfrac{M^n (M-1)}{2} \log \dfrac{t}{t_0 + 1},
\end{equation}
and:
\begin{equation}
	\label{log_cap_approx_long_correl}
	t^\star = \dfrac{M^n(M-1)}{2 \left\langle \log \dfrac{\p_{x|\mathbf{Y}}}{\R_{x|\mathbf{Y}}} \right\rangle}.
\end{equation}

\section{Laplace's rule of succession with prior information}

\subsection{The modified Laplace's rule with prior information}

In order to decrease the duration of the learning phase which is responsible for the initial loss in the log capital, a possibility is to use a non-uniform prior in the derivation of Laplace rule. 
Before starting to bet, the gambler could reasonably 
assume that the odds chosen by the bookmaker are probably not random, but contain some useful information  which can be exploited in a non-uniform prior. To do so, we use a conjugate prior which, in the case of Kelly's horse races, is the multinomial distribution:
\begin{equation}
	\label{eq:mult_prior}
	\prob{\mathbf{b}} =  \dfrac{\tau!}{a_1^{\tau} ! \cdots a_M^{\tau} !} \B_1^{a_1^{\tau}} \cdots \B_M ^{a_M^{\tau}} ~ \delta \left( \sum_x \B_x - 1 \right),
\end{equation}
where the $a_x^{\tau}$'s are free parameters and  $\tau \equiv \sum_x a_x^{\tau}$.
As both the prior and the likelihood are multinomial the two merge into a multinomial posterior: 
\begin{equation}
	\label{eq:modified_posterior}
	\condprobs{\mathbf{b}}{\mathbf{n}^t} \propto \B_1^{n_1^t + a_1^{\tau}} \cdots \B_M ^{n_M^t+ a_M^{\tau}} ~ \delta \left( \sum_x \B_x - 1 \right).
\end{equation}
Taking the expectation of the $x$-th component $\B_x$ with the measure defined by \eqref{eq:modified_posterior} gives a modified version of Laplace's rule of succession for a non-uniform prior :  
\begin{equation}
	\label{eq:pslapl_est}
	\B_x^{\text{\tiny md-LAPL}, ~ t+1} = \dfrac{\tilde{n}^{\tilde{t}}_x +1}{\tilde{t} + M},
\end{equation}
where $\tilde{t} \equiv t +\tau$ and $\tilde{n}^{\tilde{t}}_x \equiv n^t_x + a^\tau_x$.
The interpretation of \eqref{eq:pslapl_est} is that this prior is equivalent to taking into account $\tau \in \mathbb{R}$ fictional races before the first one, 
with $a_x^{\tau}$ being the equivalent of $n_x^t$ during these $\tau$ fictive races.  

Following the same route as with the uniform prior, one finds this reduced posterior \eqref{eq:mult_prior}:
\begin{equation}
	\label{eq:biased_reduced_posterior}
	\condprobs{\B_x}{\mathbf{n}^t} = \dfrac{(\tilde{t}+M-1)!}{\tilde{n}^{\tilde{t}}_x! ~ (\tilde{t} - \tilde{n}^{\tilde{t}}_x + M - 2) ! } \B_x^{\tilde{n}^{\tilde{t}}_x} (1-\B_x)^{\tilde{t} - \tilde{n}^{\tilde{t}}_x + M - 2}.
\end{equation}
At this point, the parameters of the prior (the $a_i^{\tau}$'s) need to be specified. Assuming the bookmaker is well informed of the 
real $\mathbf{p}$, the gambler should favor an initial bet distribution closer to $\mathbf{r}$ over a random distribution. 
This can be done by imposing:
\begin{equation}
	\label{eq:ps_lapl_hypothesis}
	\B_x^{\text{\tiny md-LAPL},~t=1} = \dfrac{a_x^{\tau}+1}{\tilde{t}+M} \equiv \R_x,
\end{equation}
which leads to: 
\begin{equation}
	\label{eq:constraint_prior}
	a_x^{\tau} = \R_x \cdot (\tau + M ) - 1.
\end{equation}
In order for the information in possession of the bookmaker to be exploitable, the prior should remain bounded.  
Thus, for all components $x$, the parameters of the reduced prior \eqref{eq:biased_reduced_posterior} should be strictly positive. 
This means :
\begin{equation}
	\label{eq:condition_prior}
	\forall x,~ \begin{cases} 
		a_x^{\tau} > 0 \\
		\tau - a_x^{\tau} + M - 2 > 0
	\end{cases}.
\end{equation}
Replacing $a_x^{\tau}$ by its expression given by \eqref{eq:constraint_prior} one finds that the first condition reduces to $\tau > \frac{1}{\underset{x}{\min}~\R_x}  - M $, while the second implies $\tau > \frac{1}{1 - \underset{x}{\max}~\R_x} - M $. 
Finally :
\begin{equation}
	\label{eq:constrain_t_tilde}
	\tau > \max \left[ \dfrac{1}{\underset{x}{\min}~\R_x} , ~ \dfrac{1}{1-\underset{x}{\max}~\R_x} \right] - M \geq 0.
\end{equation}
Since \eqref{eq:constraint_prior} already enforces $\sum_x a_x^{\tau} = \tau$, $\tau$ becomes the only free parameter of our problem.
Hence choosing any value of $\tau$ fulfilling \eqref{eq:constrain_t_tilde} induces a set of $a_x^{\tau}$ fulfilling \eqref{eq:ps_lapl_hypothesis}. \\
Note that a similar strategy could be built with any other form of prior information.
For instance, when this information is contained in a distribution $\boldsymbol{\pi}$, the gambler should bet 
according to : 
\begin{equation}
	\B_x^{\text{\tiny md-LAPL},~t=1} = \dfrac{a_x^{\tau}+1}{\tau+M} = \pi_x.
\end{equation}

\subsection{Limitations of the modified Laplace's strategy}

It is important to highlight that introducing a non-uniform prior only benefits the short term behavior of the log-capital, because one can show that 
the asymptotic regret of the modified Laplace's strategy is unchanged by the new choice 
of the prior: $ \left\langle  \Delta (t) \right\rangle  \sim \frac{M-1}{2} \log  t$ (see. Appendix B for the proof).  This also implies that the learning time 
$t^*$ is unchanged in the modified Laplace's strategy.

Further, it is expected that using the inverse-odds as prior information is advantageous only if the latter are more informative 
than using the uniform bet $1/M$ initially in the original Laplace's rule. 
In other words, there is the prerequisite that this inverse-odds distribution is closer to $\mathbf{p}$ than a uniform distribution would be. 
From Eq. \ref{eq:avg_capital_increases}, one can see that this means that the original strategy should start with a learning phase 
with a negative growth rate. 

If the original strategy has already a positive growth rate initially, then the bets are already better than what the odds distribution could provide, 
and in that case, the modified Laplace's strategy should not be used because it will not bring a benefit over the original Laplace's strategy. 
Indeed, using the inverse-odds as prior information results in a strategy \textit{between} 
Laplace's original strategy and the null strategy $\mathbf{b}^{\text{\tiny NULL}} = \mathbf{r}$, and the null strategy, as we mentioned earlier, has a vanishing 
capital growth rate. 
Figure \ref{fig} summarizes what we discussed, showing alongside, Laplace's strategy (orange solid curve) and the modified Laplace's based on the inverse-odds (blue solid curve) when the bookmaker has an accurate knowledge of $\mathbf{p}$. We also tested that when the original Laplace strategy has a positive slope initially, the modified Laplace's strategy can also start
with a positive slope, albeit with a smaller value, and therefore the modified Laplace strategy in this case does not
bring any benefit.

Finally, we note that the idea of using non-uniform priors has also been used in the litterature on portfolios. 
The latter is referred as \emph{$\mu$-weighted universal portfolios} \cite{cover_universal_1996}. The basic idea is 
 to replace the uniform measure, which was only enforcing the normalization of the bets in \eqref{eq:univ_estimator_general} 
 by a different arbitrary measure $\mu$, which plays the role of a prior. The authors of this reference 
used a Dirichlet distribution for $\mu$, which is a continuous measure, analogous to the multinomial prior we used.

\section{Conclusion}

In this work, we build an adaptive version of Kelly's horse race model based on Bayesian inference, and we clarify 
its connection to the theory of universal portfolios.
In practice, this strategy suffers a significant drop in the capital of the gambler up to a certain time. 
We recover the known logarithmic dependence on time for this gambler's regret, for both correlated and uncorrelated races. 
We then build an improved adaptive strategy, in which the gambler exploits the information contained in the distribution 
of the odds of the bookmaker. When this distribution is informative, the cost of the learning phase can be significantly reduced, 
without affecting the long term growth rate. 
An interesting question in this context would be to study the limit of the performance of our adaptive strategy using the notion of 
directed information  \cite{permuter_interpretations_2011,Kobayashi_2019}.

As mentioned in the introduction, one motivation for this work was to understand the process of adaptation in biological evolution and in 
particular its cost. In this correspondance between gambling and biology, the environment plays a similar role as the race results,
 phenotypic decisions taken by biological systems are the equivalent of the bets, the multiplicative rate of the biological population is the equivalent of the odds, 
 and the population growth rate is the equivalent of the long term capital growth rate. 
The adaptive strategy which we have described in this paper is a form of learning, by which individuals leverage the information  
acquired from observing past environmental states. One benefit of our work is the characterization of this initial learning phase 
with its specific time scale.

Naturally, the biological problem is vastly more complex than the model we looked at: the information acquired by biological systems from the environment 
is often partial, corrupted and distorded \cite{Rosemary_2022}. The benefit of sensing is significant 
when the environment is time-dependent and individuals have memory \cite{rivoire_value_2011}. Further, the growth of a biological population can also
act back on the environment. Some of these effects could be taken into account in our model by making odds time-dependent 
or dependent on the bets \cite{giunta_spatio_2020}, extensions which could also be interesting in the context of investment strategies. 
While adaptation in biology comes with complex costs associated with the making and the maintenance of the sensing/learning mechanisms, 
the approach of this paper provides a simple and direct way to model adaptation and its cost, which we  hope will be useful 
for future endeavours on this topic.

\section*{Acknowledgements}
We acknowledge fruitful discussions with R. Pugatch. DL acknowledges support from (ANR-11-LABX-0038, ANR-10-IDEX-0001-02).
\\

\section*{Appendix A \\ Average log-capital behavior for uncorrelated races}

Each component of $\mathbf{n}^t$ is given by the sum of independent R.V.: 
\begin{equation}
	n_x^t = \sum_{i=1}^t \left( n_x^i - n_x^{i-1} \right) \equiv \\ \sum_{i=1}^t w_x^i,
\end{equation}
where $w_x^i = 1$ if $x$ wins the race $i$ (with probability $\p_x$), $0$ if it looses (with probability $(1-\p_x)$). One immediately sees that: $\E_W \left[ w_x \right] = \p_x $ and $\sigma_x^2 \equiv \E_W \left[  \left( w_{x} - \E_W \left[ w_x \right] \right)^2 \right] = \p_x(1 - \p_x)$. Hence the application of the central limit theorem (CLT) gives:
\begin{equation}
	\label{CLT_n}
	n_x^t \underset{t \to \infty}{\approx} \p_x t + \sqrt{t} \sigma_{x} \cdot z_t,
\end{equation}
where $z_t$ is the realization at time $t$ of $Z$, a standard Gaussian R.V. of mean 0 and variance 1. Note that the leading order of \eqref{CLT_n} ensures that Laplace's rule converges to $\p_x$ when the number of race increases.
Plugging the expression of $n^t_x$ obtained with the CLT in $\log \B_{x_{t+1}}^{\text{\tiny LAPL},~ t+1}$ reads:
\begin{equation}
	\log \B_{x}^{\text{\tiny LAPL},~ t+1}  \label{eq:asympt_log_b} \approx \log \p_{x} + \dfrac{1}{\sqrt{t}}  \dfrac{\sigma_{x}}{\p_{x}}  z_{t+1} \\ -\dfrac{1}{2}\left( \dfrac{\sigma_{x}}{\p_{x}}  z_{t+1} \right)^2 + \dfrac{1}{\p_{x}t}  - \dfrac{M}{t}.
\end{equation} 
The expansion \eqref{eq:asympt_log_b} of the log will be valid when $t > \mathrm{max} \left[ \frac{1}{\underset{x}{\mathrm{min}}~\p_x} , ~ M \right] $. 
Because the sum of the components of $\mathbf{p}$ is equal to 1, $ \underset{x}{ \mathrm{min}}~\p_x \leq \frac{1}{M}$.
The condition of the expansion is then $t > \underset{x}{ \mathrm{min}}~\p_x^{-1} $. 
Hence:
{\small \begin{equation}
	\label{eq:avg_discrepancy_kelly_adapt}
	\left< \log \p_{x} - \log \B_{x}^{\text{\tiny LAPL},~ t+1} \right> \approx \left< \dfrac{\p_{x}M - 1}{\p_{x}t}  \right> \\ + \left< \dfrac{1}{2}\left( \dfrac{\sigma_{x}}{\p_{x}}  z_{t+1} \right)^2 \right> - \left< \dfrac{1}{\sqrt{t}}  \dfrac{\sigma_{x}}{\p_{x}}  z_{t+1} \right>.
\end{equation}}
The ensemble average is defined here as: $\left\langle (\cdot) \right\rangle \equiv \E_X \left[ \E_Z \left[ \left. (\cdot) \right] \right| \underline{p}   \right]  = \sum_x \p_x \cdot \E_Z \left[ (\cdot) \right] $. 
Following this definition \eqref{eq:avg_discrepancy_kelly_adapt} reduces to:
\begin{equation}
	\label{eq:avg_discrepancy_kelly_adapt_2}
	\left< \log \p_{x} - \log \B_{x}^{\text{\tiny LAPL},~ t+1} \right> \approx \dfrac{M-1}{2t}.
\end{equation}
As the regret is an additive quantity, it can be split:
\begin{equation}
	\label{eq:decomp_delta}
	\left\langle \Delta (t) \right\rangle = \left\langle \Delta (t_0)\right\rangle + \left\langle \sum_{i= t_0+1}^t \log \p_{x} - \log \B_{x}^{\text{\tiny LAPL},~ i} \right\rangle .
\end{equation}
Our previous result on the average discrepancy between Kelly's strategy \eqref{eq:avg_discrepancy_kelly_adapt_2} rests upon the expansion of the logarithm \eqref{eq:asympt_log_b}, so $t_0$ should be chosen such that the latter is valid. 
Then taking $t_0 > \underset{x}{ \mathrm{min}}~\p_x^{-1}$ one could insert \eqref{eq:avg_discrepancy_kelly_adapt_2} in \eqref{eq:decomp_delta}:
\begin{equation}
	\left\langle \Delta (t) \right\rangle \approx \left\langle \Delta (t_0)\right\rangle + \dfrac{M-1}{2} \sum_{i= t_0+1}^t \dfrac{1}{i} .
\end{equation}
The terms of the harmonic series are known to diverge in a logarithmic fashion, hence we recover the expression given in the main text:
\begin{equation}
	\label{asympt_delta}
	\left\langle  \Delta (t) \right\rangle   \approx \left\langle  \Delta (t_0) \right\rangle   + \dfrac{M-1}{2} \log \dfrac{t}{t_0+1}.
\end{equation}
Finally noticing that: 
\begin{equation}
	\label{eq:dkl_theory}
	\dkl \left( \underline{p} \Vert \mathbf{b}^{\text{\tiny LAPL}, ~ t} \right)  = \left\langle \log \p_x - \log \B_x^{\text{\tiny LAPL}, ~ t} \right>,
\end{equation} 
then \eqref{eq:avg_discrepancy_kelly_adapt_2} provides \eqref{asympt_dkl} of the main text for the KL-divergence between Laplace's estimate and $\underline{p}$. \\

\section*{Appendix B \\ Long-run equivalence between $\mathbf{b}^{\text{\tiny md-LAPL}}$ and $\mathbf{b}^{\text{\tiny LAPL}}$ \label{appen_B}}

One can proceed in the same way as in the case of pseudo-Laplace's estimate to find the discrepancy from Kelly's estimate :
\begin{equation}
	\label{eq:discrepancy_kelly_pseudo}
	\log \p_x - \log \B_x^{\text{\tiny md-LAPL},~t+1} \approx \dfrac{\tau+M}{t} + \dfrac{1}{2t}\dfrac{1-\p_x}{\p_x} z_t^2  \\ -\sqrt{\dfrac{1-\p_x}{\p_x t}} z_t  - \dfrac{ a_x^{\tau} + 1}{\p_x t}.
\end{equation}
As the average of the last term in \eqref{eq:discrepancy_kelly_pseudo} is $ \left\langle \dfrac{ a_x^{\tau} + 1}{\p_x \cdot t} \right\rangle = \sum_x \p_x \cdot \dfrac{ a_x^{\tau} + 1}{\p_x t} = \dfrac{\tau + M}{t} $, the average discrepancy from Kelly's estimate is asymptotically the same for a pseudo and a classical Laplace's estimate:
\begin{equation}
	\left\langle \log \p_x - \log \B_x^{\text{\tiny md-LAPL},~t+1} \right\rangle \approx \dfrac{M-1}{2t}.
\end{equation}
This is reasonable because in the long run the information contained in the prior becomes negligible with respect to the information contained in the likelihood.

\section*{Appendix C \\ Proof of the equivalence between $\mathbf{b}^\text{\tiny LAPL}$ and $\mathbf{b}^\text{\tiny PORTF}$}
Keeping in mind that:
\begin{equation}
	\label{eq:int_mult}
		\intsimpldetail{r}{\prod_{\tilde{x}=1}^M  \B_{\tilde{x}}^{n_{\tilde{x}}}} = \dfrac{\prod_{\tilde{x}} n_{\tilde{x}} ! }{(\sum_{\tilde{x}} n_{\tilde{x}} + M -1)!} ~ r^{\sum_{\tilde{x}} n_{\tilde{x}}+M-1},
\end{equation}
where $r \in \mathbb{R}_+$ and $\forall x,~n_x \in \mathbb{R}_+$ (for any real number $a$, we will write $a ! \equiv \Gamma (a+1)$), we can perform  an explicit computation of \eqref{eq:univ_estimator_general} with the assumption of Kelly's model: \eqref{eq:univ_estimator_general}:
\begin{equation} 
	\begin{split}
		\intsimpldetail{1}{\B_x ~ C \left[ \mathbf{b}, t  \vert \mathbf{x}^t \right] } & =  O ~ \intsimpldetail{1}{\prod_{\tilde{x}} \B_{\tilde{x}}^{n_{\tilde{x}}^t + \delta_{x,\tilde{x}}}} \\
	 & =  O ~ \dfrac{(n_x^t+1)!~\prod_{\tilde{x} \neq x } n_{\tilde{x}}^t!}{(t + M)! },
	\end{split}		 
\end{equation}
where $O \equiv \prod_{x=1}^M \odds_x^{n_x^t}$. The denominator is computed similarly:
\begin{equation}
	\intsimpldetail{1}{ C \left[ \mathbf{b}, t  \vert \mathbf{x}^t \right] } = O ~ \dfrac{\prod_{\tilde{x}} n_{\tilde{x}}^t!}{(t + M -1 )! }.
\end{equation}
Taking the quotient leads to the claimed equivalence between universal portfolio and Laplace's estimate: 
\begin{equation}
	\B^{\text{\tiny PORTF},~t+1}_x =  \dfrac{n_x^t + 1}{t + M} = \B^{\text{\tiny LAPL},~t+1}_x.
\end{equation}	

%

\section*{Appendix D \\ Average log-capital behavior for markovian races}

The CLT on the components of $\mathbf{n}^t$ should be modified to take into account correlations. 
Indeed correlations add a new random variable $\tau_y^t$, defined as the total number of time horse $y$ won during the $t$ past races. 
$\tau_y^t$ could also be expressed via the CLT and the expansion of $n_{x|y}^t$ now involves $\tau_y^t$ instead of $t$:
\begin{equation}
	\label{eq:clt_makov}
	\begin{split}
		n_{x | y }^t & \approx \p_{ x | y } \tau_y^t + \sqrt{\tau_y^t} \sigma_{x | y } \cdot z_t \\
		\tau_y^t & \approx \p^\text{s}_y t + \sqrt{t} \sigma_y \cdot w_t
	\end{split},
\end{equation}
where $\sigma_{x|y} \equiv \p_{x|y} (1 - \p_{x|y}) $ (and similarly for $\sigma_{y}$) and $w_t$, $z_t$ are two independent realizations of a standard normal random variable.
Note in addition that, as our underlying Markov chain is irreducible, the steady-state probability of winning $\mathbf{ p }^\text{s} $ is well defined.
Expanding the square root yields:
\begin{equation}
	\label{eq:tau_expansion}
	\sqrt{\tau_y^t } \approx \sqrt{ \p^\text{s}_y t} + \dfrac{1}{2} \dfrac{\sigma_y w_t}{\sqrt{\p^\text{s}_y}}.
\end{equation}
The logarithm of our estimate is given by: 
\begin{equation}
	\label{eq:log_bet_markov} \log \B_{x | y}^{\text{\tiny LAPL},~ t+1} = \log \left(  n_{x | y}^{t} + 1 \right) - \log \left( \tau_{y}^{t} + M \right) .
\end{equation}
In what comes next we will denote $(\diamond)$ terms that are of null average.
The first terms of \eqref{eq:log_bet_markov} reads: 
\begin{equation}
	\label{eq:markov_lapl_log_num}
	\log \left(  n_{x | y}^{t} + 1 \right) \approx \log \p_{x | y} + \log \p^\text{s}_{y} t + \dfrac{1}{\p^\text{s}_{y} \p_{x | y} t} \\ - \dfrac{1}{2 t} \left( \dfrac{\sigma_{y}^2 w_{t+1}^2}{\p^\text{s}_{y} } + \dfrac{\sigma_{x | y}^2 z_{t+1}^2}{\p_{x | y}^2 \p^\text{s}_{y}} \right) +  (\diamond).
\end{equation}
While the expansion of the second term of \eqref{eq:log_bet_markov} yields:
\begin{equation}
	\label{eq:markov_lapl_log_denom}
	\log \left( \tau_{y}^{t} + M \right) \approx \log \p^\text{s}_{y} t + \dfrac{M}{\p^\text{s}_{y} t} - \dfrac{1}{2} \dfrac{\sigma_{y}^2 w_{t+1}^2}{\sqrt{t} \p^\text{s}_{y}} + (\diamond).
\end{equation}
The expansions \eqref{eq:markov_lapl_log_num} \eqref{eq:markov_lapl_log_denom} are valid provided $t$ follows $t > \mathrm{max} \left[ t_1, t_2 \right]$ with: 
\begin{equation}
\label{eq:condition_t0}
\begin{array}{l}
	t_1 \equiv \underset{x, y}{\mathrm{max}} \left[ \dfrac{1}{\p^\text{s}_y \p_{ x | y }} \right] \\
	t_2 \equiv \underset{x,y}{\mathrm{max}} \left[ ~ \dfrac{1}{2}\dfrac{1}{\p^\text{s}_y}\sqrt{\dfrac{1-\p^\text{s}_y}{\p^\text{s}_y}}\sqrt{\dfrac{1-\p_{x|y}}{\p_{x|y}}} ~ \right] 
\end{array}.
\end{equation}
Note that the condition to expand $\sqrt{\tau_y^t}$ in the CLT of $n_{x|y}^t$ are automatically satisfied for any $t$ fulfilling \eqref{eq:condition_t0}. \\
The ensemble average in the case of correlated races is defined by: $\left\langle  (\cdot) \right\rangle \equiv \sum_{x,y} \p_{x,y} \cdot \E_{Z,W} \left[ (\cdot) \right] =  \sum_{x,y} \p^\text{s}_y \p_{x|y} \cdot \E_{Z,W} \left[ (\cdot) \right]$.
Applying this definition reads: 
\begin{equation}
	\label{eq:avg_discrepancy_kelly_markov}
	\left\langle  \log \p_{x | y} - \log \B_{x | y}^{\text{\tiny LAPL},~ t+1} \right\rangle \approx  \dfrac{M(M-1)}{2 t}.
\end{equation}
Summing \eqref{eq:avg_discrepancy_kelly_markov} from a $t_0$ that satisfies the condition  \eqref{eq:condition_t0} gives: 
\begin{equation}
	\label{asympt_delta_markov}
	\left\langle \Delta (t) \right\rangle  \approx \left\langle \Delta (t_0) \right\rangle  + \dfrac{M(M-1)}{2} \log \dfrac{t}{t_0 + 1}.
\end{equation}

\section*{Appendix E\\ Arbitrary long correlations}

The case of longer correlations is similar to classical Markov races. 
Indeed let's call $n$ the length of the correlation, the realization at time $t$ of the R.V. $X$ depends now on $\mathbf{Y}_{t} \equiv \left[ x_{t}, x_{t-1}, \cdots, x_{t-n+1} \right] $. 
As for $t \geq n$, $\mathbb{P} \left[  x_{t+1} \left| \mathbf{Y}_{t} \mathbf{Y}_{t-1} \cdots \mathbf{Y}_{n} \right. \right] = \mathbb{P} \left[  x_{t+1} \left| \mathbf{Y}_{t} \right. \right] \equiv \p_{x_{t+1} | \mathbf{Y}_t}$, hence we recover a Markovian process. 
Then the CLT is expressed as before:
\begin{equation}
	\label{eq:clt_correl}
	\begin{split}
		n_{x | \mathbf{Y} }^t & \approx \p_{ x | \mathbf{Y} } \tau_{\mathbf{Y}}^t + \sqrt{\tau_{\mathbf{Y}}^t} \sigma_{x | \mathbf{Y} } \cdot z_t \\
		\tau_{\mathbf{Y}}^t & \approx \p^\text{s}_{\mathbf{Y}} t + \sqrt{t} \sigma_{\mathbf{Y}} \cdot w_t
	\end{split}.
\end{equation}
Doing the exact same steps as above one finds that :  
\begin{equation}
	\log \p_{x | \mathbf{Y}} - \log \B_{x | \mathbf{Y}}^{\text{\tiny LAPL},~ t+1} \approx \dfrac{1}{ \p_{x | \mathbf{Y}  } ~ \p_{\mathbf{Y}}^\text{s} t} -\dfrac{M}{\p_{\mathbf{Y}}^\text{s} t} \\ - \dfrac{1}{2 t}  \dfrac{\sigma_{x | \mathbf{Y}}^2 ~ z_{t+1}^2}{ \p_{x | \mathbf{Y}}^2 ~ \p_{\mathbf{Y}}^\text{s}} + (\diamond),
\end{equation}
again $(\diamond)$ denotes terms that are of null average. 
The ensemble average is defined as : $ \left\langle  (\cdot) \right\rangle \equiv   \sum_{x, \mathbf{Y}} \p^\text{s}_{\mathbf{Y}} ~ \p_{x| \mathbf{Y}} \cdot \E_{Z,W} \left[ (\cdot) \right] $.
Applying this definition, remembering that: $ \sum_{x, \mathbf{Y}}~1 = M^{n+1}$ and $ \sum_{x, \mathbf{Y}} \p_{ x | \mathbf{Y} } = M^n$, reads:
\begin{equation}
	\label{eq:avg_discrepancy_kelly_correl}
	\left\langle  \log \p_{x | \mathbf{Y}} - \log \B_{x | \mathbf{Y}}^{\text{\tiny LAPL},~ t+1} \right\rangle \approx \dfrac{M^n(M-1)}{2t}.
\end{equation}
Summing \eqref{eq:avg_discrepancy_kelly_correl} yields the expression for the regret for arbitrary correlated races:
\begin{equation}
	\label{eq:avg_delta_correl}
	\left\langle \Delta (t) \right\rangle = \left\langle \Delta (t_0) \right\rangle + \dfrac{M^n(M-1)}{2} \log \dfrac{t}{t_0 + 1},
\end{equation}
which is valid for a $t_0$ following the same constraint \eqref{eq:condition_t0} replacing $\p_{x|y}$ and $\p_{y}^\text{s}$ by $\p_{ x | \mathbf{Y}}$ and $\p_{\mathbf{Y}}^\text{s}$ respectively.

\section*{References}

\providecommand{\newblock}{}

\end{document}